\documentclass[11pt,prd,onecolumn,amsmath,amssymb,aps,floats,floatfix, nofootinbib]{revtex4-2}

\usepackage[colorlinks=true,urlcolor=blue,anchorcolor=blue,citecolor=blue,filecolor=blue,linkcolor=blue,menucolor=blue,linktocpage=true]{hyperref} % should be commented out if the tex file will be compiled with latex in arXiv!!! (pdflatex is fine)

\usepackage[inline]{enumitem}
\usepackage[multidot]{grffile}  % allow the name of figures to include dots
\usepackage{dcolumn}
\usepackage{bm}
\usepackage{amsmath}
\usepackage{amsfonts}
\usepackage{amssymb}
\usepackage{color}
\usepackage[table]{xcolor}
\usepackage{float}
\usepackage{latexsym}
\usepackage{slashed} % slash mark
\usepackage{pstricks}
\usepackage{indentfirst}
\usepackage{mathrsfs}
\usepackage{multirow}
\usepackage{epsfig,psfrag}
\usepackage{graphicx}
\usepackage{enumitem}
\usepackage{geometry,amssymb,yfonts}
\usepackage{yhmath}
\usepackage{anysize}
\usepackage{subfigure}
\usepackage{mathtools}
\usepackage{setspace} % spacing
\usepackage[utf8]{inputenc} % accept utf-8 input encoding
\usepackage[scientific-notation=true]{siunitx} % comprehensive units
\usepackage{feynmp-auto}
\usepackage{verbatim}

\usepackage[normalem]{ulem}
\usepackage{longtable}

\graphicspath{{fig/}}

\allowdisplaybreaks % allow eqnarray breaks
\newcommand{\Uone}{\mathrm{U}(1)}
\newcommand{\UoneX}{\mathrm{U}(1)_\mathrm{X}}

\begin{document}

\title{Interpreting the KM3-230213A PeV Neutrino Event via Vector Dark Matter Decay and Its Multi-Messenger Signatures}

\author{Yu-Hang Su$^{a}$}
\author{Si-Yu Chen$^{a}$}
\author{Chengfeng Cai$^{b}$}\email[Corresponding author. ]{caichf3@mail.sysu.edu.cn}
\author{Hong-Hao Zhang$^{a}$}\email[Corresponding author. ]{zhh98@mail.sysu.edu.cn}

\affiliation{$^a$School of Physics, Sun Yat-sen University, Guangzhou 510275, China}
\affiliation{$^b$School of Science, Sun Yat-Sen University, Shenzhen 518107, China}

\begin{abstract}
The KM3NeT Collaboration recently reported the detection of an ultra-high-energy neutrino event KM3-230213A with a reconstructed energy of $220^{+570}_{-110}$ PeV, the most energetic astrophysical neutrino ever detected. The absence of convincing electromagnetic counterparts motivates exploration of exotic origins beyond standard astrophysical processes. We present a vector dark matter model based on a new $\UoneX$ gauge symmetry to interpret this event through superheavy dark matter decay. Our analysis demonstrates that dark matter lifetimes in the range $7.3 \times 10^{28}$–-$2.9 \times 10^{30}$ s can successfully account for the KM3-230213A event while satisfying stringent constraints from gamma-ray observations. Moreover, the spontaneous breaking of $\UoneX$ in our model naturally predicts cosmic string formation, generating a stochastic gravitational wave background with string tension $4.5 \times 10^{-11} \lesssim G\mu \lesssim 1.2 \times 10^{-10}$, consistent with recent pulsar timing array observations. This multi-messenger consistency across neutrinos, gamma-rays, and gravitational waves validates our vector dark matter interpretation of the KM3-230213A event while providing testable predictions for upcoming multi-wavelength experiments.

\end{abstract}

\maketitle
\tableofcontents
%\clearpage

\section{Introduction}

The KM3NeT Collaboration recently reported the detection of an ultra-high-energy (UHE) neutrino event, named as KM3-230213A, with a reconstructed energy of $220^{+570}_{-110}$~PeV from the direction RA = $94.3^\circ$ and Dec = $-7.8^\circ$~\cite{KM3NeT:2025npi}. This event represents the most energetic astrophysical neutrino ever detected, exceeding previous records by more than an order of magnitude.

Several hypotheses have been proposed to explain the origin of KM3-230213A~\cite{Boccia:2025hpm, Fang:2025nzg, Dzhatdoev:2025sdi, Brdar:2025azm, Dvali:2025ktz, Klipfel:2025jql, Narita:2025udw, Muzio:2025gbr, Satunin:2025uui, Alves:2025xul, Yang:2025kfr, Khan:2025gxs, Kohri:2025bsn, Jiang:2025blz}. Conventional astrophysical scenarios attribute such UHE neutrinos to cosmic ray interactions ($pp$ or $p\gamma$ processes) in powerful accelerators like blazars or gamma-ray bursts. However, producing a 220 PeV neutrino requires proton acceleration to EeV energies—beyond the known capabilities of Galactic or nearby extragalactic sources. Comprehensive multi-wavelength follow-up observations identified 17 potential blazar counterparts within the $99\%$ confidence region but found no compelling electromagnetic associations, despite hints of temporal correlations in some candidates~\cite{KM3NeT:2025bxl, KM3NeT:2025aps, KM3NeT:2025vut}.

The absence of a convincing astrophysical counterpart motivates exploration of exotic origins beyond standard astrophysical processes. One compelling possibility is the decay of superheavy dark matter (DM), which could naturally produce neutrinos at these extreme energies without requiring conventional cosmic accelerators. In the standard cosmological model, DM constitutes about $26\%$ of the total energy density of the universe~\cite{Planck:2018vyg}, yet its particle nature remains unknown. Superheavy DM is regarded as one of the promising candidates, especially as TeV-scale Weakly Interacting Massive Particles (WIMPs) face increasingly stringent constraints from direct detection experiments~\cite{XENON:2023cxc, PandaX:2024qfu, LZ:2024zvo}. Unlike normal WIMPs that achieve relic abundance through thermal freeze-out~\cite{Lee:1977ua}, superheavy DM may originate from non-thermal freeze-in mechanisms due to the perturbative unitarity bound~\cite{Asaka:2005cn, Asaka:2006fs,Hall:2009bx}. Such particles can produce a sharp energy cutoff in the neutrino spectrum, a feature that could explain the KM3-230213A event and be tested by future neutrino observations. Importantly, the decay of superheavy DM is expected to generate not only neutrinos but also high-energy gamma-rays through electroweak radiation and subsequent hadronic cascades. These gamma-rays, after propagation and attenuation via interactions with the interstellar radiation field (ISRF), could be observed by experiments like LHAASO~\cite{LHAASO:2024upb, LHAASO:2024lnz}, CASA-MIA~\cite{CASA-MIA:1997tns} and PAO~\cite{Castellina:2019huz}. Therefore, any DM interpretation of the KM3-230213A event must be consistent with stringent constraints from both neutrino and gamma-ray observations, typically requiring DM lifetimes far beyond the age of the universe.

Since no viable particle within the Standard Model (SM) can play the role of DM, new physics beyond the SM is required. The $\Uone$ extension is one of the simplest and most natural way to introduce a dark sector for DM, making it a widely studied in many new physics models~\cite{Lao:2020inc, Su:2024oml, Su:2025mxv, Shan:2023uii, Qiu:2023wbs, Liu:2022evb}. In particular, if the $\Uone$ symmetry spontaneously breaks in the early universe, it can produce a kind of topological defect known as cosmic strings. After formation, cosmic strings quickly enter a scaling regime, continuously producing loops through the intersection and reconnection of long strings. Oscillations of these loops emit stochastic gravitational waves (GWs) lasting until the present day, which can be probed by current and future GW interferometers such as EPTA~\cite{EPTA:2023sfo, EPTA:2023xxk}, NANOGrav~\cite{NANOGrav:2023gor, NANOGrav:2023hvm}, and LISA~\cite{2017arXiv170200786A}. Therefore, a $\Uone$-extended DM model not only provides a natural origin for the UHE neutrino flux observed in the KM3-230213A event but also predicts a GW spectrum that can be tested against existing observational constraints, offering a multi-messenger approach to probe this interesting DM scenario.

In this work, we present a vector DM model based on a new $\UoneX$ gauge symmetry. The model introduces three new particles charged under $\UoneX$: two Weyl fermions, $\eta$ and $\zeta$, and a complex scalar $\Phi$. To generate the observed light neutrino masses, we also include two singlet right-handed neutrinos, $N_1$ and $N_2$, which constitute the minimal setup required to reproduce the masses of all three generations of light neutrinos. The $\UoneX$ symmetry is spontaneously broken when $\Phi$ acquires a nonzero vacuum expectation value $v_\Phi$. Consequently, Yukawa interactions involving $\eta$, $\zeta$, $N_1$, $N_2$, and $\Phi$ generate mass mixing between the $\UoneX$ charged fermions and right-handed neutrinos, thereby enabling the DM to decay into light neutrinos. We emphasize that our model simultaneously achieves several key objectives:
\begin{enumerate}
\item Successfully generates light neutrino masses consistent with experimental data via the seesaw mechanism.

\item Provides a viable explanation for the KM3-230213A event through new Yukawa interactions that allow DM to decay into light neutrinos, while satisfying the stringent constraints on DM lifetime.

\item Achieves the correct relic density through the freeze-in process $\phi \to XX$, where $\phi$ is the CP-even component of $\Phi$.

\item  Successfully produces a stochastic GW background via cosmic strings generated by the spontaneous breaking of $\UoneX$ in the early universe, consistent with current GW observational constraints.
\end{enumerate}
Remarkably, our model accomplishes this diverse and stringent set of experimental requirements with only six independent parameters. The predicted fluxes of UHE neutrino, gamma-ray, and the GW spectrum offer exciting targets for current and upcoming experiments.

This paper is organized as follows. In Sect.~\ref{Dark Matter Model}, we introduce our $\UoneX$-extended vector DM model, focusing on the neutrino and scalar sectors. In Sect.~\ref{Dark Matter Phenomenology}, we explore the phenomenological implications of the model, including the DM relic density and neutrino/gamma-ray fluxes from DM decay. In particular, we identify viable parameter regions that can account for the KM3-230213A event. Sect.~\ref{Cosmic string and Gravitational waves} presents the predicted GW spectrum induced by cosmic strings, along with the allowed parameter space testable by future GW experiments. Finally, we conclude in Sect.~\ref{Conclusions}.

\section{A Vector Dark Matter Model}
\label{Dark Matter Model}

In this section, we provide a detailed description of our vector DM model based on a new gauge symmetry $\UoneX$. The corresponding gauge field $ X_\mu$ serves as our DM candidate. This model introduces three new particles charged under $\UoneX$: two Weyl fermions, $\eta$ and $\zeta$, with charges $+1$ and $-1$ respectively, and one complex scalar $\Phi$ with charge $+1$. Such a charge assignment ensures the theory remains free of gauge anomalies. Moreover, to generate the correct light neutrino masses, two singlet right-handed neutrinos $N_1$ and $N_2$ are also included. The model can be regarded as an extension of the minimal seesaw scenario~\cite{King:1999mb, King:2002nf, Frampton:2002qc}.

\subsection{Neutrino sector}
\label{Neutrino sector}

The Beyond-the-Standard-Model (BSM) renormalizable and gauge-invariant Lagrangian for our model is given by,
\begin{eqnarray}\label{Lagrangian}
\mathcal{L}_{\text{BSM}} &=& -\frac{1}{4} X_{\mu \nu} X^{\mu \nu} + \eta^\dagger i\bar{\sigma}^\mu D_\mu \eta + \zeta^\dagger i\bar{\sigma}^\mu D_\mu \zeta -m_{\eta\zeta}(\eta \zeta + \text{h.c.}) + N_I^\dagger i\bar{\sigma}^\mu \partial_\mu N_I - \frac{1}{2}(M_{IJ} N_I N_J \nonumber\\
&+& \text{h.c.}) -\sqrt{2}(y_{I\zeta} \Phi \zeta N_I + y_{I\eta} \Phi^\dagger \eta N_I + \text{h.c.}) + (D^\mu \Phi)^\dagger D_\mu \Phi - V(\Phi) + \mathcal{L}_{\text{seesaw}}(N_1,N_2),\nonumber\\
\end{eqnarray}
where $m_{\eta \zeta}$ denotes the Dirac mass of the four-component Dirac field $(\eta,\zeta^\dagger)$, and $M_{IJ}$ corresponds to the Majorana mass matrix of the fields $N_1$ and $N_2$. The Yukawa couplings $y_{I\eta(\zeta)}$ characterize the gauge-invariant interaction strength between $\Phi$, $\eta(\zeta)$, and $N_I$. Furthermore, $V(\Phi, H)$ denotes the scalar potential involving $\Phi$ and the SM Higgs field $H$, while $\mathcal{L}_{\text{seesaw}}(N_1, N_2)$ represents the effective Lagrangian after electroweak symmetry breaking for the seesaw sector, which is given by,
\begin{eqnarray}
\mathcal{L}_{\text{seesaw}}(N_1,N_2) = N_1 (d \nu_{eL} + e \nu_{\mu L} + f \nu_{\tau L}) + N_2 (a \nu_{eL} + b \nu_{\mu L} + c \nu_{\tau L}) + \text{h.c.},
\end{eqnarray}
where $a,b,c,d,e,f$ denote the Dirac mass between $N_{1,2}$ and the SM left-handed neutrinos in the flavor basis. 

After the spontaneous breaking of $\UoneX$ symmetry, $\Phi$ acquires a vacuum expectation value (VEV) $v_\Phi$. In the unitary gauge, $\Phi$ can be expressed as,
\begin{eqnarray}
\Phi = \frac{1}{\sqrt{2}} (v_\Phi+\phi),
\end{eqnarray}
where $\phi$ is a new CP-even Higgs-like particle. The DM $X$ acquires a mass given by,
\begin{eqnarray}\label{DM mass}
m_{X} = g_X v_\Phi,
\end{eqnarray}
where $g_X$ denotes the gauge coupling of $\UoneX$. The Majorana mass matrix in the basis of $\eta$, $\zeta$, $N_1$ and $N_2$ can be expressed as,
\begin{eqnarray}
(M_N)_{(\eta,\zeta,N_1,N_2)} =
\begin{pmatrix}
0 & m & y_{1\eta} v_\phi &  y_{2\eta} v_\phi\\
m & 0 & y_{1\zeta} v_\phi &  y_{2\zeta} v_\phi\\
y_{1\eta} v_\phi & y_{1\zeta} v_\phi & M_{11} & M_{12} \\
y_{2\eta} v_\phi & y_{2\zeta} v_\phi & M_{21} & M_{22} 
\end{pmatrix}.
\end{eqnarray}
For simplicity, we consider the minimal setup where $M_{12} = y_{1\eta} =y_{2\eta}  = y_{2\zeta} = 0$, i.e., only $\zeta$ couples to $N_1$ directly. The mass matrix for $\eta$, $\zeta$ and $N_1$ then simplifies to be
\begin{eqnarray}
(M_N)_{(\eta,\zeta,N_1)} = 
\begin{pmatrix}
0 & m_{\eta\zeta} & 0 \\
m_{\eta\zeta} & 0 & M_{1\zeta}\\
0 & M_{1\zeta} & M_{11} \\
\end{pmatrix},
\end{eqnarray}
where we have defined $M_{1\zeta} = y_{1\zeta} v_\phi$. We can then construct the complete mass matrix for all relevant Weyl fermions ($\eta$, $\zeta$, $N_1$, $N_2$, $\nu_{eL}$, $\nu_{\mu L}$, $\nu_{\tau L}$) in our model as,
\begin{eqnarray}
M_N =
\begin{pmatrix}
0 & m_{\eta\zeta} & 0 & 0 & 0 & 0 & 0\\
m_{\eta\zeta} & 0 & M_{1\zeta} & 0 & 0 & 0 & 0\\
0 & M_{1\zeta} & M_{11} & 0 & d & e & f\\
0 & 0 & 0 & M_{22} & a & b & c \\
0 & 0 & d & a & 0 & 0 & 0 \\
0 & 0 & e & b & 0 & 0 & 0 \\
0 & 0 & f & c & 0 & 0 & 0 \\
\end{pmatrix}.
\end{eqnarray}
In this work, we only focus on the hierarchical case, which requires: $a,b,c,d,e,f \ll M_{1\zeta} \ll M_{11}, M_{22}, m_{\eta\zeta}$. As a result, $M_N$ can be diagonalized through a two-step procedure: first diagonalizing the $3 \times 3$ submatrix $(M_N)_{(\eta,\zeta, N_1)}$, and then the full $7 \times 7$ matrix $M_N'$. 

The submatrix $(M_N)_{(\eta, \zeta, N_1)}$ can be diagonalized by a unitary matrix $P$ as: $P^T M_N P = \text{diag}(\lambda_1,\lambda_2,\lambda_3)$. Using perturbation theory, the eigenvalues $\lambda_i$ and the rotation matrix $P$ can be approximated to first order in the small parameter $M_{1\zeta}/m_{\eta\zeta}$ as,
\begin{eqnarray}
&& \lambda_1 = -m_{\eta\zeta}, \quad \lambda_2 = m_{\eta\zeta}, \quad m_3 = M_{11}, \\
\nonumber \\
P &&=
\begin{pmatrix}
\frac{1}{\sqrt{2}} & \frac{1}{\sqrt{2}} & \frac{M_{1\zeta} m_{\eta\zeta}}{M_{11}^2-m_{\eta\zeta}^2} \\
-\frac{1}{\sqrt{2}} & \frac{1}{\sqrt{2}} & \frac{M_{1\zeta} M_{11}}{M_{11}^2-m_{\eta\zeta}^2}\\
\frac{M_{1\zeta}}{\sqrt{2}(m_{\eta\zeta}+M_{11})} & \frac{M_{1\zeta}}{\sqrt{2}(m_{\eta\zeta}-M_{11})} & 1 \\
\end{pmatrix}.
\end{eqnarray}
Note that the eigenvalues coincide with their leading-order values where $M_{1\zeta} = 0$, as there are no corrections at first-order perturbation. We then define a new basis via the transformation: $(\eta, \zeta, N_1)^T = P(\eta', \zeta', N_1')^T$. In the full basis ($\eta'$, $\zeta'$, $N_1'$, $N_2$, $\nu_{eL}$, $\nu_{\mu L}$, $\nu_{\tau L}$), the complete mass matrix $M_N'$ can be approximated as follows:
\begin{eqnarray}
M_N' =
\begin{pmatrix}
-m_{\eta\zeta} & 0 & 0 & 0 & \epsilon' d & \epsilon' e & \epsilon' f\\
0 & m_{\eta\zeta} & 0 & 0 & \epsilon' d & \epsilon' e & \epsilon' f\\
0 & 0 & M_{11} & 0 & d & e& f\\
0 & 0 & 0 & M_{22} & a & b & c \\
\epsilon' d & \epsilon' d & d & a & 0 & 0 & 0 \\
\epsilon' e & \epsilon' e & e & b & 0 & 0 & 0 \\
\epsilon' f & \epsilon' f & f & c & 0 & 0 & 0 \\
\end{pmatrix},
\end{eqnarray}
where $\epsilon' \equiv M_{1\zeta}/(\sqrt{2}m_{\eta\zeta})  \ll 1$, and $m_{\eta\zeta} \gg M_{11}$ is assumed. To leading order in $\epsilon'$, the off-diagonal elements connecting $\eta'$ and $\zeta'$ to the other states are zero, so we need only consider the mass matrix in the sub-basis $(N_1', N_2, \nu_{eL}, \nu_{\mu L}, \nu_{\tau L})$, which reduces to the conventional seesaw case:
\begin{eqnarray}
(M_N')_{(N_1', N_2, \nu_{eL}, \nu_{\mu L}, \nu_{\tau L})} =
\begin{pmatrix}
M_R & M_\nu \\
M_\nu^T	 & 0
\end{pmatrix}.
\end{eqnarray}
After diagonalization, the eigenvalues of $N_1'$ and $N_2$ are nearly unchanged, while the effective mass matrix of the light neutrinos is given by,
\begin{eqnarray}
m_\nu &=& M_\nu^T M_R^{-1} M_\nu \nonumber\\
&=&
\frac{1}{M_{11}}\begin{pmatrix}
d^2 & de & df \\
de & e^2 & ef \\
df & ef & f^2 
\end{pmatrix}
+
\frac{1}{M_{22}}\begin{pmatrix}
a^2 & ab & ac \\
ab & b^2 & bc \\
ac & bc & c^2 
\end{pmatrix},
\end{eqnarray}
where we have absorbed the overall unphysical minus sign into the fermion field. Consequently, another rotation is thus needed to diagonalize the matrix $m_\nu$ and obtain the light neutrino mass eigenstates $(\nu_1, \nu_2, \nu_3)$. In this work, we adopt the framework of sequential dominance (SD) to naturally generate a hierarchical neutrino mass spectrum: $m_1 \ll m_2 \ll m_3$~\cite{King:1999mb, King:2002nf, King:1998jw, King:1999cm}. In this scenario, each column of the Dirac mass matrix predominantly corresponds to a specific physical neutrino mass, with one right-handed neutrino controlling the heaviest mass, the second dominating the next heaviest, and the third effectively decoupled, yielding the lightest mass. The SD framework imposes the following hierarchical conditions:
\begin{eqnarray}
\frac{d^2,e^2,f^2}{M_{11}} \gg \frac{a^2,b^2,c^2}{M_{22}}.
\end{eqnarray}
Moreover, motivated by the large atmospheric angle $\theta_{23}$, large solar angle $\theta_{12}$ and small reactor angle $\theta_{13}$, we adopt the following parameter relations: $d=0$, $e=f$, $b=na$, $c=(n-2)a$~\footnote{Phases are ignored here for simplicity. A complete treatment including phases can be found in Ref.~\cite{King:2002nf}.}, known as constrained sequential dominance (CSD) for a real parameter $n$~\cite{King:2005bj, Antusch:2011ic, King:2013iva, King:2013xba, King:2013hoa, Bjorkeroth:2014vha}. Taking $n=3$ as a representative example, the light neutrino masses are approximately given by,
\begin{eqnarray}
m_{1} \simeq 0, \quad m_{2} \simeq \frac{3a^2}{M_{22}}, \quad m_{3} \simeq \frac{2e^2}{M_{11}}.
\end{eqnarray}
Under the CSD conditions, once the Majorana masses of $N_1$ and $N_2$ are specified, the Dirac masses $a$ and $e$ can be uniquely determined from the corresponding observed light neutrino masses. For a normal neutrino mass hierarchy, the neutrino masses are $m_2 \simeq 0.0086$~eV and $m_3 \simeq 0.05$~eV~\cite{Esteban:2024eli}. When $M_{11} = 10^{10}$ GeV, we have $e = 0.5$ GeV. The $5 \times 5$ rotation matrix $U$ diagonalizing $(M_N')_{(N_1', N_2, \nu_{eL}, \nu_{\mu L}, \nu_{\tau L})}$ can be written as follows:
\begin{eqnarray}
U =
\begin{pmatrix}
1-\frac{1}{2}\theta \theta^\dagger & \theta \\
-\theta^\dagger	 & 1-\frac{1}{2}\theta^\dagger\theta
\end{pmatrix},
\end{eqnarray}
where $\theta_{i\nu}=M_{\nu}/M_{ii}$. After the second rotation $U$, the transformed basis becomes: ($\eta''$, $\zeta''$, $N_1''$, $N_2'$, $\nu_{eL}'$, $\nu_{\mu L}'$, $\nu_{\tau L}'$). Since $\theta$ is extremely small, we have the relations $1-\frac{1}{2}\theta \theta^\dagger \simeq 1$ and $1-\frac{1}{2}\theta^\dagger\theta\simeq 1$, which allow us to approximate $(\nu_{eL}', \nu_{\mu L}', \nu_{\tau L}') \simeq (\nu_{eL}, \nu_{\mu L}, \nu_{\tau L})$. 

To the first order of $\epsilon'$, mass eigenvalues retain their leading-order values, whereas the rotation matrix receives first-order corrections. The full $7 \times 7$ matrix $U$ thus takes following form:
\begin{eqnarray}
U
=
\begin{pmatrix}
1 & 0 & 0 & 0 & \epsilon' \frac{d}{m_{\eta\zeta}} & \epsilon' \frac{e}{m_{\eta\zeta}} & \epsilon' \frac{f}{m_{\eta\zeta}}\\
0 & 1 & 0 & 0 & -\epsilon' \frac{d}{m_{\eta\zeta}} & -\epsilon' \frac{e}{m_{\eta\zeta}} & -\epsilon' \frac{f}{m_{\eta\zeta}}\\
0 &0 & 1 & 0 & \frac{d}{M_{11}} & \frac{e}{M_{11}} & \frac{f}{M_{11}}\\
0 & 0 & 0 & 1 & \frac{a}{M_{22}} & \frac{b}{M_{22}} & \frac{c}{M_{22}}\\
-\epsilon' \frac{d}{m_{\eta\zeta}} & \epsilon' \frac{d}{m_{\eta\zeta}} & -\frac{d}{M_{11}} & -\frac{a}{M_{22}} & 1 & 0 & 0\\
-\epsilon' \frac{e}{m_{\eta\zeta}}  & \epsilon' \frac{e}{m_{\eta\zeta}} & -\frac{e}{M_{11}} & -\frac{b}{M_{22}} & 0 & 1 & 0\\
-\epsilon' \frac{f}{m_{\eta\zeta}} & \epsilon' \frac{f}{m_{\eta\zeta}} & -\frac{f}{M_{11}} & -\frac{c}{M_{22}} & 0 & 0 & 1\\
\end{pmatrix}.
\end{eqnarray}
Consequently, the complete rotation matrix relating the original gauge basis to the mass basis, ($\eta''$, $\zeta''$, $N_1''$, $N_2'$, $\nu_{eL}'$, $\nu_{\mu L}'$, $\nu_{\tau L}'$), is then given by the product $V = P U$, which is explicitly expressed as follows,
\begin{eqnarray}
V = 
\begin{pmatrix}
\frac{1}{\sqrt{2}} & \frac{1}{\sqrt{2}} & -\epsilon'  & 0 & -\epsilon' \frac{d}{M_{11}} & -\epsilon' \frac{e}{M_{11}} & -\epsilon' \frac{f}{M_{11}}\\
-\frac{1}{\sqrt{2}} & \frac{1}{\sqrt{2}} & -\epsilon' \frac{M_{11}}{m} & 0 & -(\sqrt{2}+1)\epsilon'\frac{d}{m_{\eta\zeta}} & -(\sqrt{2}+1)\epsilon'\frac{e}{m_{\eta\zeta}} & -(\sqrt{2}+1)\epsilon'\frac{f}{m_{\eta\zeta}}\\
\epsilon' & \epsilon' & 1 & 0 & \frac{d}{M_{11}} & \frac{e}{M_{11}} & \frac{f}{M_{11}}\\
0 & 0 & 0 & 1 & \frac{a}{M_{22}} & \frac{b}{M_{22}} & \frac{c}{M_{22}} \\
-\epsilon' \frac{d}{m_{\eta\zeta}} & \epsilon'\frac{d}{m_{\eta\zeta}} & -\frac{d}{M_{11}} &  -\frac{a}{M_{22}} & 1 & 0 & 0 \\
-\epsilon' \frac{e}{m_{\eta\zeta}} & \epsilon'\frac{e}{m_{\eta\zeta}} & -\frac{e}{M_{11}} &  -\frac{b}{M_{22}} & 0 & 1 & 0 \\
-\epsilon' \frac{f}{m_{\eta\zeta}} & \epsilon'\frac{f}{m_{\eta\zeta}} & -\frac{f}{M_{11}} &  -\frac{c}{M_{22}} & 0 & 0 & 1 \\
\end{pmatrix}.
\end{eqnarray}

The DM particle $X$ decays via the channels: $X \to \nu_{\mu L} \nu_{\mu L}, \nu_{\tau L} \nu_{\tau L}, \nu_{\mu L} \nu_{\tau L}$. Here, we assume that DM is much lighter than all heavy neutrinos and the scalar $\phi$, so any decay process involving $\eta''$, $\zeta''$, $N_1''$, $N_2'$, and $\phi$ are kinetically forbidden. By imposing the CSD conditions, the squared scattering amplitude for the process $X \to \nu_I \nu_J$ can be approximated as,
\begin{eqnarray}
|\mathcal{M}|^2_{X \to \nu_I \nu_J}
\simeq \frac{8g_X^2 \epsilon'^4 e^4 m_X^2}{S_f M_{11}^4},
\end{eqnarray}
where $S_f$ is the symmetry factor for identical particles in the final state. Finally, the total decay width of DM is given by the sum of the partial widths:
\begin{eqnarray}\label{GammaX}
\Gamma_{X} &=& \Gamma_{X\to \nu_{\mu L} \nu_{\mu L}} + \Gamma_{X\to \nu_{\tau L} \nu_{\tau L}} + \Gamma_{X\to \nu_{\mu L} \nu_{\tau L}} \nonumber\\
&=& \frac{1}{3\pi} \frac{g_X^2 \epsilon'^4 e^4 m_{X}}{M_{11}^4}.
\end{eqnarray}

\subsection{Scalar sector}
\label{Scalar sector}

The scalar potential $V(\Phi, H)$ in the lagrangian~\eqref{Lagrangian} can be written as,
\begin{eqnarray}
V(\Phi) = -\mu_\Phi^2 |\Phi|^2 + \frac{1}{2} \lambda_\Phi |\Phi|^4 -\mu_H^2 |H|^2 + \frac{1}{2} \lambda_H |H|^4 + \lambda_{H\Phi} |H|^2 |\Phi|^2.
\end{eqnarray}
After imposing the vacuum extremum conditions, the mass-squared matrix for the two CP-even scalars $(\phi, h)$ takes the form,
\begin{eqnarray}
M_E^2 = 
\begin{pmatrix}
\lambda_H v_H^2 & \lambda_{H\Phi} v_H v_\Phi \\
\lambda_{H\Phi} v_H v_\Phi & \lambda_\Phi v_\Phi^2
\end{pmatrix}.
\end{eqnarray}
The matrix can be diagonalized by an orthogonal transformation such that: $O^T M'_E O = \text{diag}(m_{h_1}^2, m_{h_2}^2)$, where the rotation matrix $O$ is given by
\begin{eqnarray}
O &=& 
\begin{pmatrix}
c_\alpha & s_\alpha \\
-s_\alpha & c_\alpha
\end{pmatrix},\\
\alpha &\simeq& \frac{\lambda_{H\Phi} v_H}{\lambda_\Phi v_\Phi}.
\end{eqnarray}
Since the DM in our model is super-heavy, requiring $v_\Phi \gg v_H$. Without assuming any hierarchy between $\lambda_{H\Phi}$ and $\lambda_\Phi$, this implies $\alpha \ll 1$, allowing us to safely treat $\phi$ and $h$ as their respective mass eigenstates.

For superheavy DM, thermal freeze-out production is not viable, as the coupling required to achieve the appropriate annihilation cross section would exceed the unitarity bound. Consequently, this motivates us to consider non-thermal freeze-in as the dominant production mechanism. In our model, the associate production process is $\phi \to X X$, with the decay width given by,
\begin{eqnarray}
\Gamma_{\phi \to XX} = \frac{3g_X^4 v_\Phi^2}{8\pi m_\phi}.
\end{eqnarray}
The DM relic density can be approximated as~\cite{Hall:2009bx},
\begin{eqnarray}\label{relic}
\Omega_X h^2 &\simeq& \frac{1.09 \times 10^{27}}{g_{*,s} \sqrt{g_{*,\rho}}} \frac{2m_{X}\Gamma_{\phi \to 2X}}{m_\phi^2} \nonumber\\
&\simeq& 0.12 \times \left(\frac{3.2 \times 10^{13}~\text{GeV}}{v_\Phi}\right)^5 \frac{1}{\lambda_\Phi^{3/2}}.
\end{eqnarray}
In summary, our model involves six parameters: $M_{11}$, $e$, $\lambda_\Phi$($m_\phi$), $v_\Phi$, $g_X$, and $\epsilon'$. These parameters are subject to four key constraints: the heaviest left-handed neutrino mass, and the mass, lifetime, relic density of the DM particle. As a result, only two parameters can be chosen as free.

\section{Phenomenology from Dark Matter Decay}
\label{Dark Matter Phenomenology}

In this section, we analyze the DM phenomenology in our model, focusing specifically on the neutrino and gamma-ray signatures arising from super-heavy DM decay, a focus motivated by the recent KM3-230213A event.

The total neutrino flux produced by DM decay consists of two components: galactic and extragalactic contributions. For the galactic component, the differential neutrino flux within an observed solid angle $\Delta \Omega$ is given by,
\begin{eqnarray}\label{flux1}
\frac{d^2\Phi^G_\nu}{dE_\nu d\Omega} = \frac{D}{4\pi m_{\text{DM}}} \sum_i \Gamma_i \frac{dN_\nu^i}{dE_\nu},
\end{eqnarray}
where $\Gamma_i$ is the partial decay width for channel $i$. In our model, these channels correspond to the processes $X \to \nu_{\mu L} \nu_{\mu L}$, $\nu_{\tau L} \nu_{\tau L}$, $\nu_{\mu L} \nu_{\tau L}$. The neutrino energy spectrum per DM decay, $dN_\nu^i/dE_\nu$, is calculated using the HDMSpectra package~\cite{Bauer:2020jay}. The astrophysical D-factor encodes the DM density distribution along the line of sight, defined as,
\begin{eqnarray}
D = \frac{1}{\Delta \Omega}\int_{\Delta \Omega} d\Omega \int_0^{s_\text{max}} ds \, \rho_{\text{DM}}\left( r\left( s,b,l \right) \right),
\end{eqnarray}
where $s$ denotes the line-of-sight distance from Earth, while $r$ denotes the radial distance from the Galactic Center (GC). These coordinates are related through:
\begin{eqnarray}
r = \sqrt{s^2 + r_\odot^2 - 2sr_\odot\cos b \cos l},
\end{eqnarray}
with $r_\odot = 8.3$ kpc being the Earth-GC distance, and $(b,l)$ being the galactic coordinates. In this work, we adopt the Navarro-Frenk-White (NFW) density profile~\cite{Navarro:1996gj}, which takes the form:
\begin{eqnarray}
 \rho_{\text{DM}}(r) = \frac{\rho_0 r_s^3}{r(r_s + r)^2},
\end{eqnarray}
where we adopt the characteristic DM density $\rho_0 = 0.318~\mathrm{GeV~cm^{-3}}$~\cite{HAWC:2017udy} and the scale radius $r_s = 20$~kpc. It is worth noting that different choices of the DM profile mainly affect the DM density near the GC~\cite{Cirelli:2010xx}. Therefore, for events located far from the GC, such as the KM3-230213A, the results are not expected to vary significantly with the choice of DM profile.

For the extragalactic component, the differential neutrino flux can be derived considering the cosmological redshift effect:
\begin{eqnarray}\label{flux2}
\frac{d^2\Phi^{EG}_\nu}{dE_\nu d\Omega} = \frac{1}{\tau_\chi 4\pi M_\chi} \int_0^{\infty} dz \frac{\rho_0 c/H_0}{\sqrt{\Omega_m(1+z)^3 + \Omega_\Lambda} (1+z)} \left(\frac{dN_\nu}{dE'_\nu}\right)\bigg|_{E'_\nu=(1+z)E_\nu},
\end{eqnarray}
where $\rho_0 = 1.15 \times 10^{-6}~\mathrm{GeVcm^{-3}}$ is the average cosmological DM density at the present-day, $c/H_0 = 1.37 \times 10^{28}$~cm is the Hubble length, and $\Omega_m = 0.315$, $\Omega_\Lambda = 0.685$ are the matter and dark energy density parameters, respectively~\cite{Planck:2018vyg}. 

\begin{figure}[!h]
\centering
\includegraphics[width=0.7\textwidth]{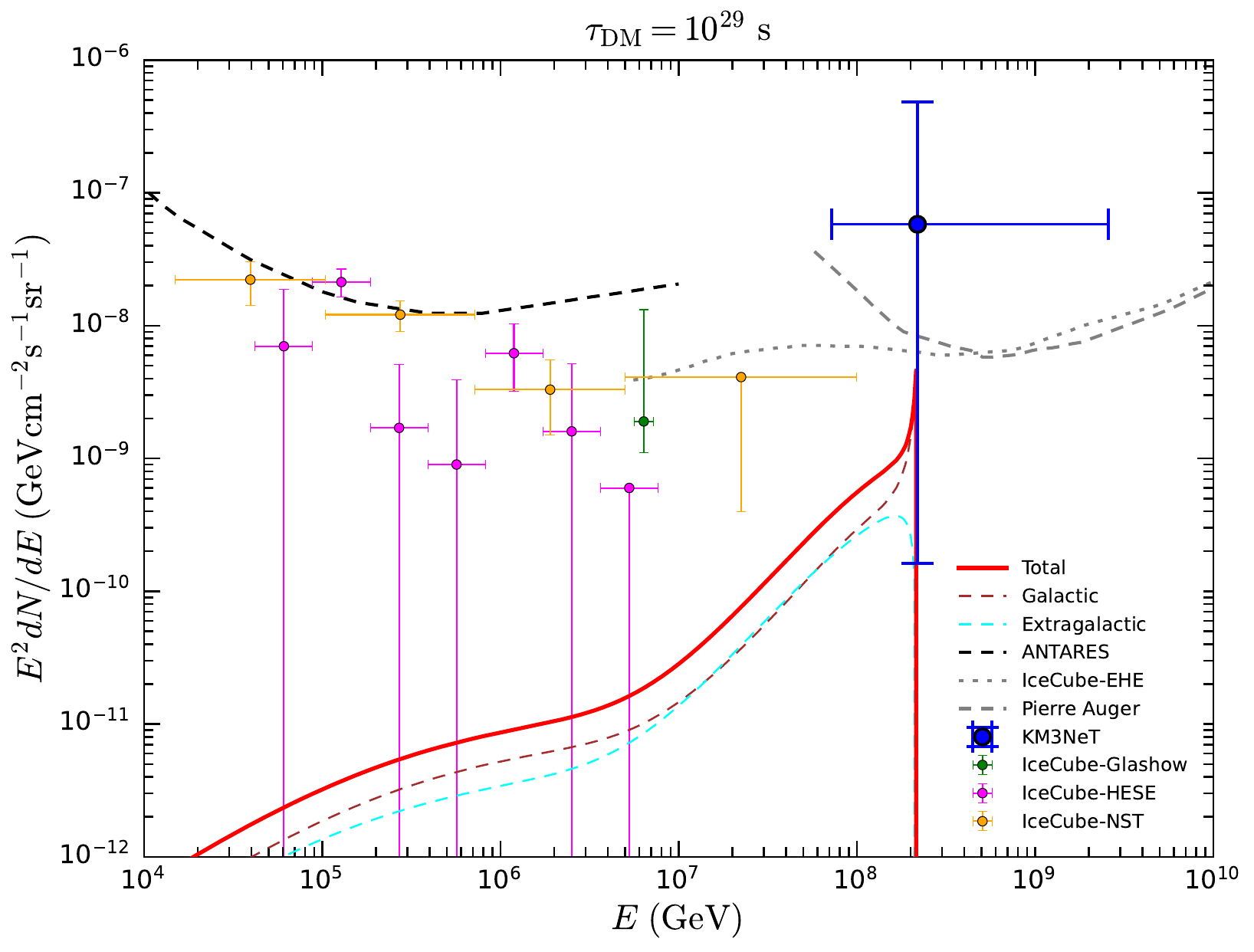}
\caption{Predicted neutrino flux from 440~PeV DM $X$ with a lifetime of $10^{29}$~s. Red (solid), brown (dashed), and cyan (dashed) lines represent total, galactic, and extragalactic contributions, respectively. Data points are shown as crosses in different colors: KM3-230213A (blue), IceCube-HESE (magenta), IceCube-NST (orange), and IceCube-Glashow (green). Upper limits are shown for ANTARES (black dashed), Pierre Auger Observatory (gray dashed), and IceCube (gray dotted).}
\label{Fig1}
\end{figure}

In Fig.~\ref{Fig1}, we show the predicted neutrino flux from our model. The red (solid), brown (dashed), and cyan (dashed) lines correspond to the total flux, galactic contribution, and extragalactic contribution, respectively. The DM mass is fixed at $440$~PeV and its lifetime is set to be $\tau_{\text{DM}}=10^{29}$~s. The flux is calculated within an angular uncertainty of $\pm 1.5^\circ$ at the $1\sigma$ confidence level (C.L.) around the arrival direction of the KM3-230213A event, which is located at RA: $94.3^\circ$ and Dec: $-7.8^\circ$. Furthermore, we assume an equal neutrino flavor ratio of $1:1:1$ at Earth, consistent with the assumptions of the KM3NeT Collaboration.
The data points are shown as follows: the blue cross marks the KM3-230213A event at the $3\sigma$ C.L.; magenta and orange crosses represent measurements from IceCube-HESE~\cite{IceCube:2020wum} and IceCube-NST~\cite{Abbasi:2021qfz}, respectively; the green cross denotes the IceCube Glashow resonance event~\cite{IceCube:2021rpz}. The upper limits are shown as black dashed lines for ANTARES ($95\%$ C.L.)~\cite{ANTARES:2024ihw}, gray dashed lines for Pierre Auger Observatory ($90\%$ C.L.)~\cite{Salamida:2023fmk}, and gray dotted lines for IceCube ($90\%$ C.L.)~\cite{IceCube:2018fhm}. Our model can account for the KM3-230213A event while remaining consistent with the upper limits.

\begin{figure}[h!]
\centering
\subfigure[\label{Fig2-1}]
{\includegraphics[width=0.48\textwidth]{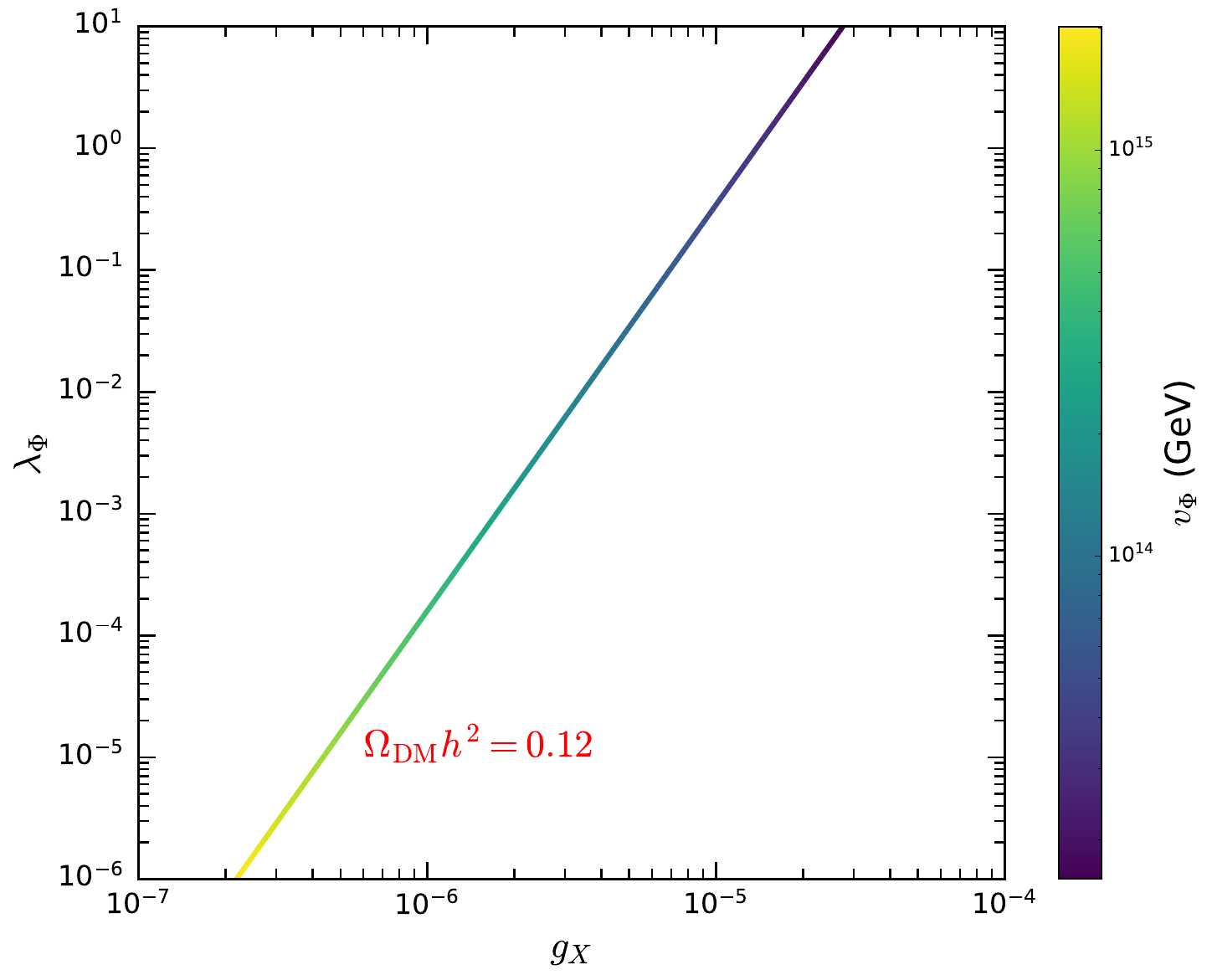}}
\hspace{.01\textwidth}
\subfigure[\label{Fig2-2}]
{\includegraphics[width=0.48\textwidth]{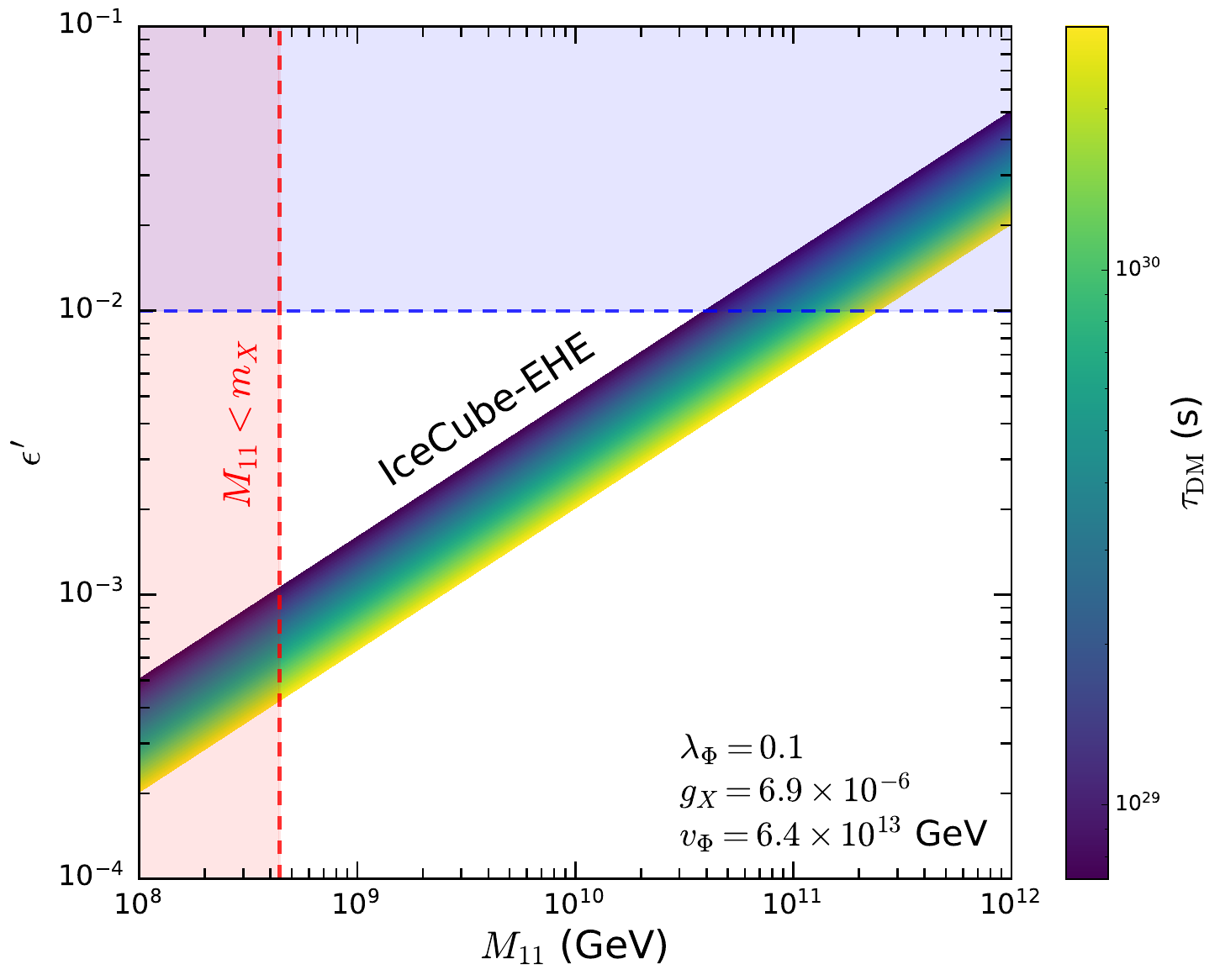}}
\hspace{.01\textwidth}
\subfigure[\label{Fig2-3}]
{\includegraphics[width=0.48\textwidth]{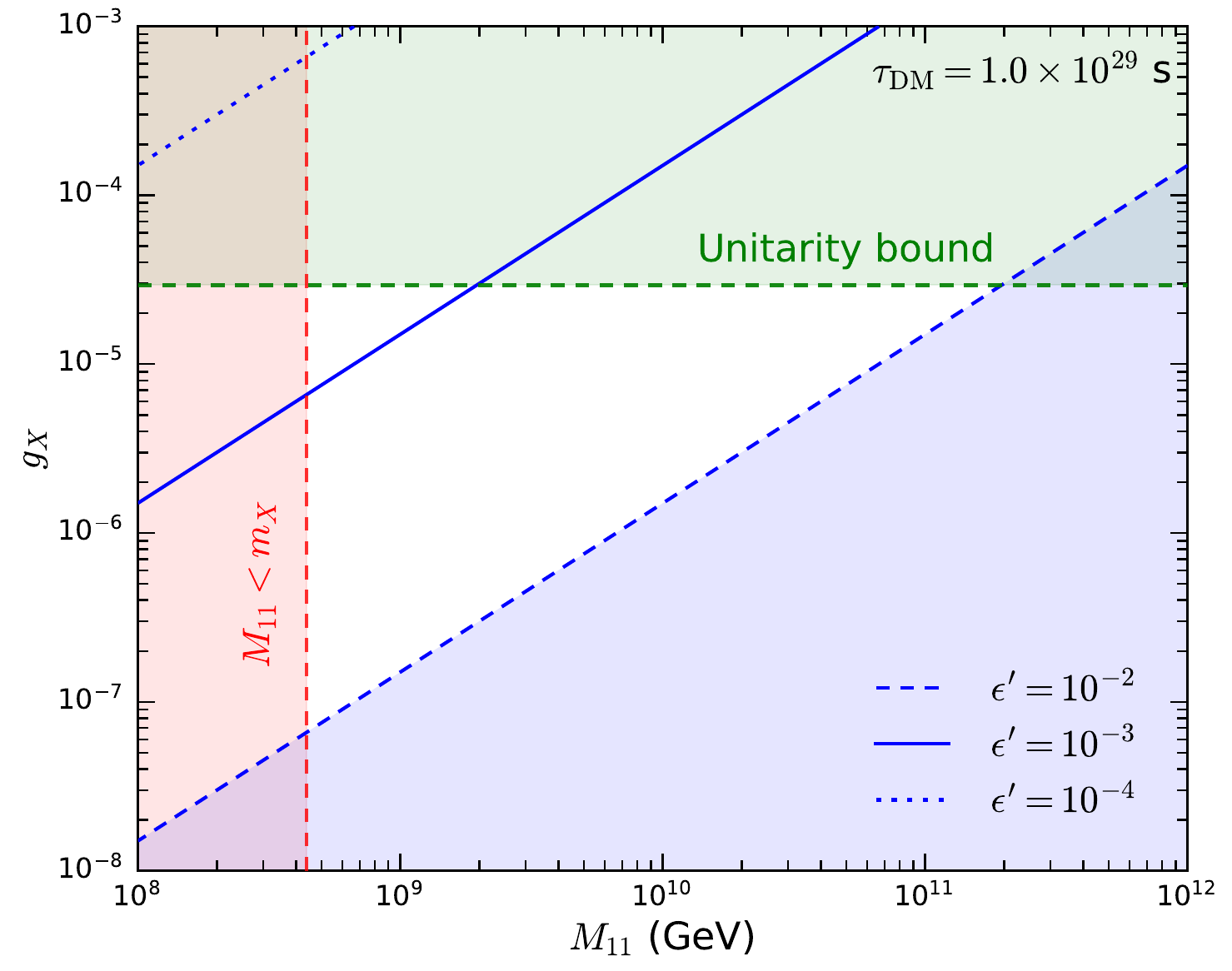}}
\caption{Parameter spaces in the $g_X$-$\lambda_\Phi$ (a), $M_{11}$-$\epsilon'$ (b), and $M_{11}$-$g_X$ (c) planes. Panel (a) shows parameter correlations for the observed DM relic density, with colors representing $v_\Phi$. Panel (b) displays the viable region for $\lambda_\Phi = 0.1$, colored by DM lifetime $\tau_{\text{DM}}$, while Panel (c) presents the parameter space for fixed $\tau_\text{DM} = 10^{29}$~s. Excluded regions are shown in red ($m_{x} < m_{N_1''}$), blue (beyond our approximation $M_{1\zeta} \ll M_{11} \ll m_{\eta\zeta}$), and green (violating unitarity, $\lambda_\Phi > 4\pi$).}
\label{Fig2}
\end{figure}

In Fig.~\ref{Fig2}, we further examine the parameter correlations in our model under various constraints. The contour in Fig.~\ref{Fig2-1} representing the observed DM relic density $\Omega_\text{DM} h^2 = 0.12$ shows a positive correlation between $g_X$ and $\lambda_\Phi$, with the color gradient indicating a negative correlation between $g_X$ and $v_\Phi$. These correlations follow directly from the DM mass expression in Eq.~\eqref{DM mass} and the relic density calculation in Eq.~\eqref{relic}.
Taking $\lambda_\Phi = 0.1$ as a benchmark (corresponding to $g_X = 6.9 \times 10^{-6}$ and $v_\Phi = 6.4 \times 10^{13}$~GeV), we examine the viable parameter space in the $M_{11}$-$\epsilon'$ plane shown in Fig.~\ref{Fig2-2}. The allowed parameter region corresponds to a neutrino flux that exhibits a peak within the $3\sigma$ confidence interval of KM3-230213A, as shown in Fig.~\ref{Fig1}, while remaining below the IceCube-EHE upper limit. This region is color-mapped according to the DM lifetime $\tau_{\text{DM}}$, which is thus constrained to the range $7.3 \times 10^{28}$~s to $2.9 \times 10^{30}$~s.
In addition, the red region representing $M_{11} < m_{X}$ (equivalent to $m_{X} > m_{N_1''}$) which is not considered in this work, while the blue region represents $\epsilon' > 10^{-2}$ is excluded due to the invalidity of the approximation $M_{1\zeta} \ll M_{11} \ll m_{\eta\zeta}$ (equivalent to $\epsilon' \ll M_{11}/m_{\eta\zeta} \ll 1$). The viable ranges for $M_{11}$ and $\epsilon'$, which determine $\tau_{\text{DM}}$ via Eq.~\eqref{GammaX}, are thus $4.4 \times 10^{8}~\mathrm{GeV} < M_{11} < 2.5 \times 10^{11}~\mathrm{GeV}$ and $4 \times 10^{-4} < \epsilon' < 10^{-2}$.
In Fig.~\ref{Fig2-3}, we turn to fix $\tau_\text{DM} = 10^{29}$~s to investigate the viable parameter space in the $M_{11}$-$g_X$ plane. The red and blue regions remain excluded following the same criteria as in Fig.~\ref{Fig2-2}, while the green region corresponds to $\lambda_\Phi < 4\pi$ is excluded due to violation of the naive unitarity bound. Solid and dotted lines represent $\epsilon' = 10^{-3}$ and $10^{-4}$, respectively. For $M_{11}$ and $g_X$, the viable ranges are thus $4.4 \times 10^{8}~\mathrm{GeV} < M_{11} < 2 \times 10^{11}~\mathrm{GeV}$ and $6.4 \times 10^{-8} < g_X < 3 \times 10^{-5}$.

\begin{figure}[!h]
\centering
\includegraphics[width=0.6\textwidth]{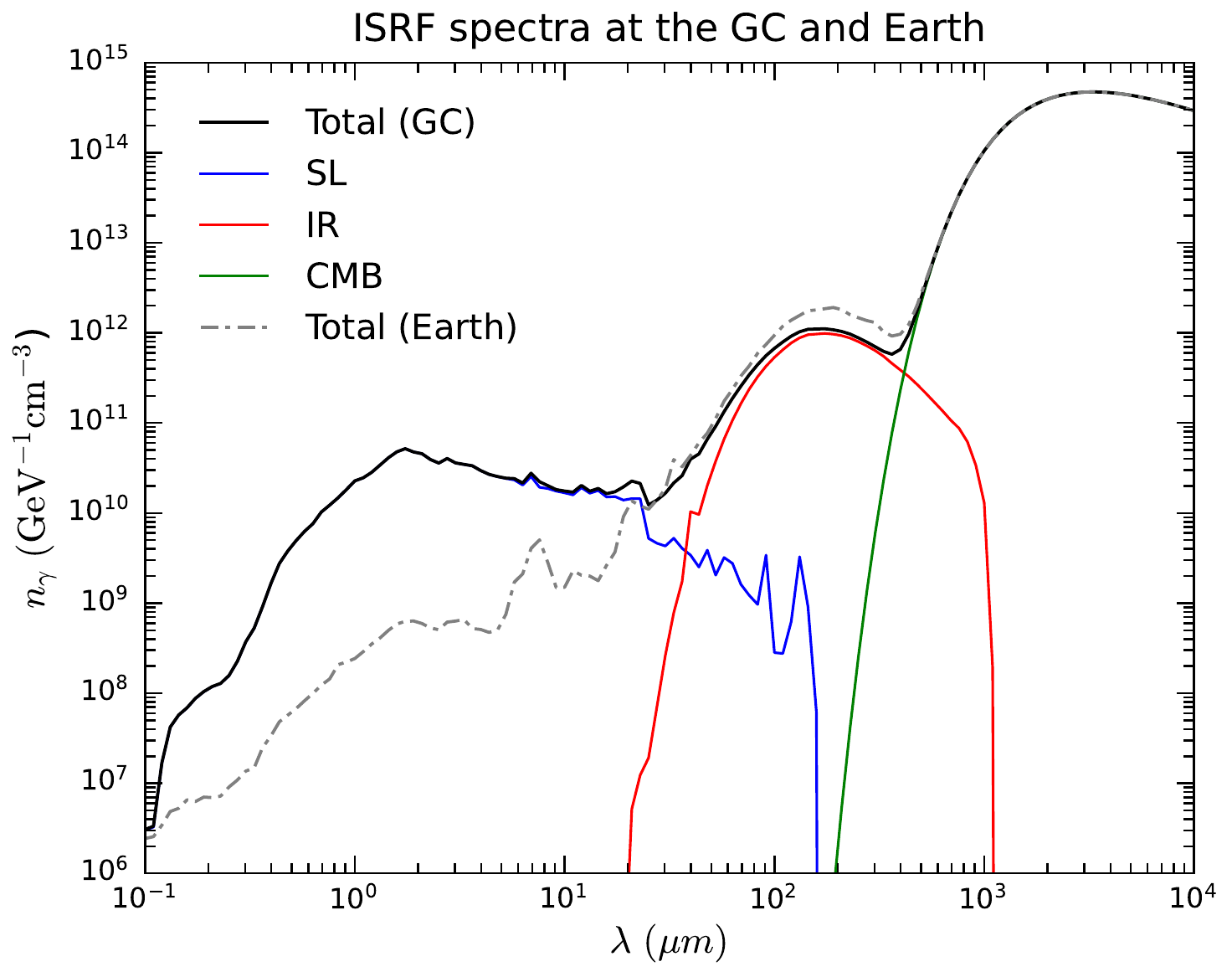}
\caption{ISRF spectrum versus photon wavelength $\lambda$ at the GC (black) and Earth (gray), with blue, red, and green curves representing SL, IR and CMB components, respectively. }
\label{Fig3}
\end{figure}

On the other hand, although gamma-rays are not produced directly from DM decay in our model, they can be generated through electroweak radiation processes. The differential gamma-ray flux can be evaluated using the same formalism as in Eqs.~\eqref{flux1} and \eqref{flux2}, by substituting the neutrino energy spectrum with the gamma-ray spectrum $dN_\gamma^i/dE_\gamma$. However, unlike the neutrino flux, the gamma-ray flux around the PeV scale suffers significant attenuation during propagation from its production site to Earth. This attenuation occurs because gamma-rays interact with the interstellar radiation field (ISRF) via the $\gamma \gamma \to e^- e^+$ pair production process. Fig.~\ref{Fig3} illustrates the full ISRF spectrum based on the R12 spatial model at the Galactic Center (GC, black curve) and Earth (gray curve), which is extracted from GALPROP~\cite{Porter:2021tlr, Porter:2017vaa, Robitaille:2012kg}. Different components dominate the spectrum at different wavelengths ($\lambda$): starlight (SL, blue), infrared radiation (IR, red), and cosmic microwave background (CMB, green). As expected, the CMB component exhibits spatial homogeneity and isotropy, resulting in identical spectra at the GC and Earth, while the SL and IR spectra are position-dependent.

\begin{figure}[!h]
\centering
\includegraphics[width=0.6\textwidth]{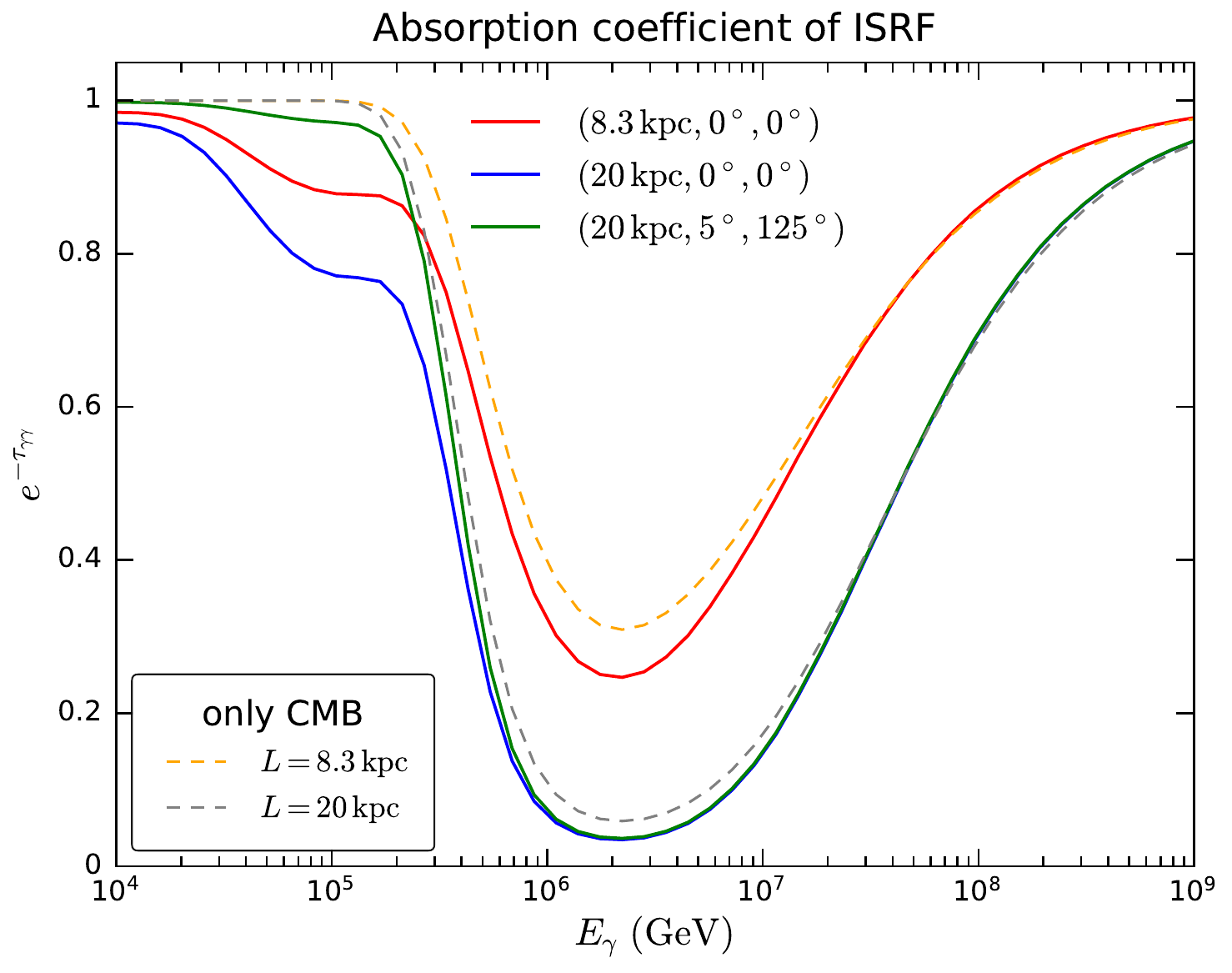}
\caption{ISRF absorption coefficient $e^{-\tau_{\gamma \gamma}(E_\gamma,L,b,l)}$ versus gamma-ray energy for three galactic positions: ($8.3~\mathrm{kpc}, 0^\circ, 0^\circ$, red), ($20~\mathrm{kpc}, 0^\circ, 0^\circ$, blue), and ($20~\mathrm{kpc}, 5^\circ, 125^\circ$, green). Orange and gray dashed lines show CMB-only absorption for $L = 8.3$ and $20~\mathrm{kpc}$, respectively.}
\label{Fig4}
\end{figure}

The optical depth for gamma-rays of energy $E_\gamma$ produced at distance $L$ is thus given by~\cite{Esmaili:2015xpa}:
\begin{eqnarray}\label{tau_gamma}
\tau_{\gamma \gamma}(E_\gamma,L,b,l) = \int_0^L dx \iint \sigma_{\gamma \gamma}(E_\gamma, \epsilon) n_\gamma(\epsilon,x,b,l) \frac{1-\cos \theta}{2} \sin \theta d\theta d\epsilon,
\end{eqnarray}
where $\sigma_{\gamma \gamma}$ is the pair production cross section:
\begin{eqnarray}
\sigma_{\gamma \gamma}(E_\gamma, \epsilon) = \frac{\pi}{2} \frac{\alpha^2}{m_e^2}(1-\beta^2) \left[ (3-\beta^4)\ln \left( \frac{1+\beta}{1-\beta} - 2\beta(2-\beta^2) \right) \right],
\end{eqnarray}
with $\alpha$ being the fine-structure constant, $m_e$ representing the the electron mass, $\theta$ indicating the angle between the interacting photons, and $\beta$ is defined as:
\begin{eqnarray}
\beta = \sqrt{1-\frac{1}{s}}, \quad s=\frac{\epsilon E_\gamma}{2m_e^2}(1-\cos \theta).
\end{eqnarray}
For the CMB component, the optical depth can be simplified by substituting the $2.73$~K blackbody spectrum into Eq.~\eqref{tau_gamma}:
\begin{eqnarray}
\tau^\text{CMB}_{\gamma \gamma}(E_\gamma, L) = \frac{-4T_\text{CMB}L}{\pi^2 E_\gamma^2} \int_{m_e}^\infty \epsilon_c^3 \sigma_{\gamma \gamma}(\epsilon_c) \ln\left[1 - e^{\frac{-\epsilon_c^2}{E_\gamma T_\text{CMB}}} \right] d\epsilon_c,
\end{eqnarray}
where $\epsilon_c = \sqrt{\epsilon E_\gamma(1-\cos \theta)/2}$. Including attenuation effects, the differential gamma-ray flux from galactic DM decay becomes:
\begin{eqnarray}\label{gaammaflux}
\frac{d^2\Phi_\gamma}{dE_\gamma d\Omega} = \frac{1}{4\pi m_{\text{DM}} } \sum_i \Gamma_i\frac{dN_\gamma^i}{dE_\gamma} \int_0^\infty ds \rho_\text{DM}(s,b,l) e^{-\tau_{\gamma \gamma}(E_\gamma,L,b,l)}. 
\end{eqnarray}
Note that we neglect the extragalactic contribution since it is strongly suppressed by pair production processes during propagation over cosmological distances~\cite{Ishiwata:2019aet}. Fig.~\ref{Fig4} shows the ISRF absorption coefficient $e^{-\tau_{\gamma \gamma}(E_\gamma,L,b,l)}$ as a function of $E_\gamma$ at three galactic positions ($L, b, l$): ($8.3~\mathrm{kpc}, 0^\circ, 0^\circ$), ($20~\mathrm{kpc}, 0^\circ, 0^\circ$), and ($20~\mathrm{kpc}, 5^\circ, 125^\circ$), shown in red, blue, and green, respectively. For comparison, we also show the absorption coefficients from the CMB component alone, $e^{-\tau^\text{CMB}_{\gamma \gamma}(E\gamma, L)}$, for $L = 8.3$~kpc and $20$~kpc as orange and gray dashed lines, respectively. The results demonstrate that the CMB dominates the attenuation above $300$~TeV, while the combined SL + IR contribution becomes significant in the $10$-$300$~TeV range. For the red and blue curves, which represent the same direction but different distances, the blue curve lies consistently below the red curve due to stronger attenuation over the longer propagation distance of $L = 20$~kpc. The green curve, despite having the same distance of $L = 20$~kpc, shows reduced SL+IR absorption due to its direction away from the GC, while maintaining the same CMB contribution as the blue curve.

\begin{figure}[!h]
\centering
\includegraphics[width=0.7\textwidth]{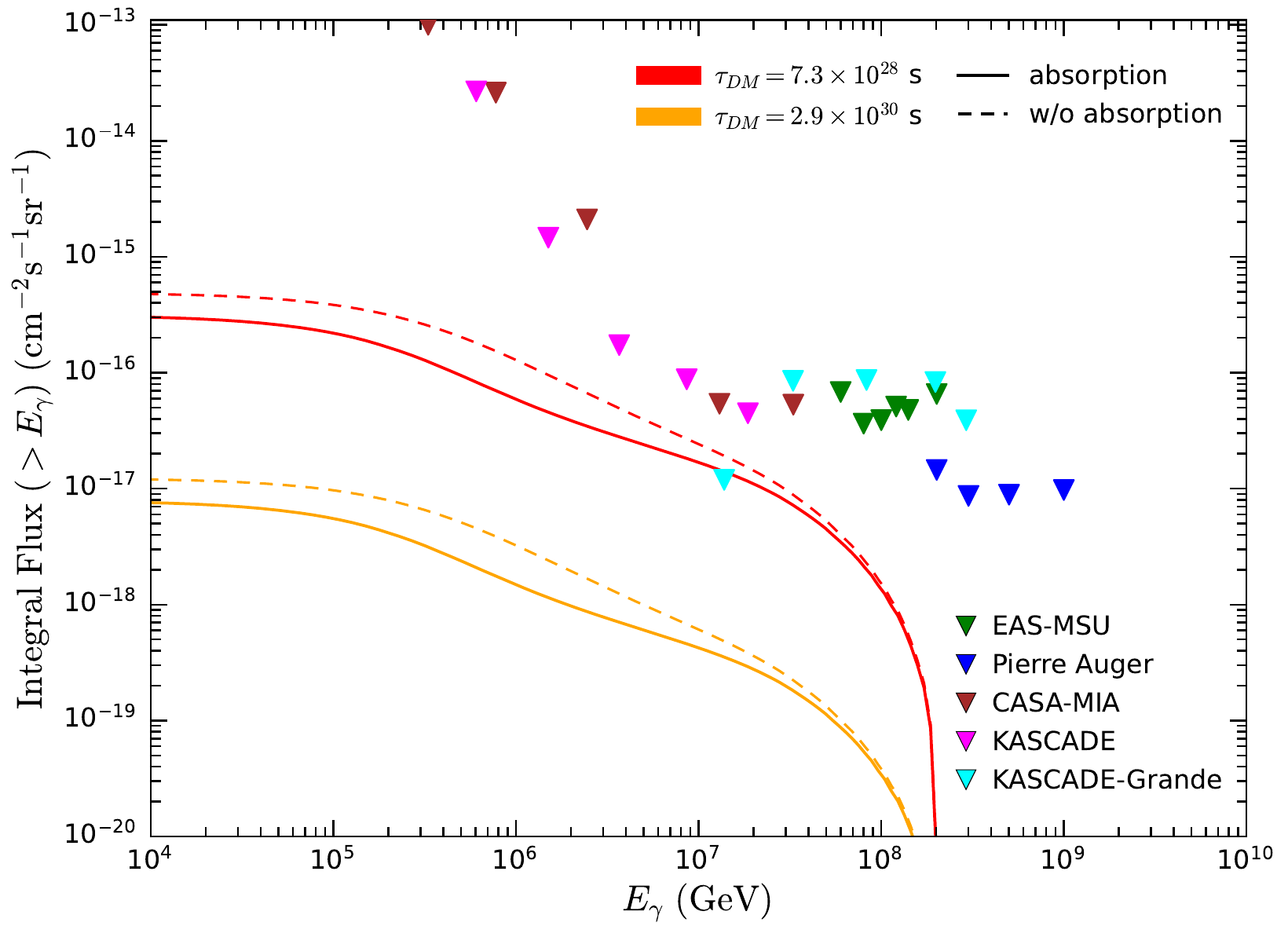}
\caption{Integral gamma-ray flux for two representative DM lifetimes: $7.3 \times 10^{28}$~s (red) and $2.9 \times 10^{30}$~s (orange). Solid and dashed curves show results with and without ISRF attenuation, respectively. Colored triangles indicate experimental upper limits from EAS-MSU (green), Pierre Auger Observatory (blue), CASA-MIA (brown), KASCADE (magenta), and KASCADE-Grande (cyan).}
\label{Fig5}
\end{figure}

To enable direct comparison with current limits on diffuse ultra-high-energy (UHE) gamma-ray flux, we calculate the integral flux averaged over all directions $b$ and $l$~\cite{Chianese:2021jke}:
\begin{eqnarray}
\Phi_\gamma(E_\gamma) = \frac{1}{4 \pi} \int_{E_\gamma}^\infty dE'_\gamma \int_{4\pi} d\Omega \frac{d^2\Phi_\gamma}{dE'_\gamma d\Omega}.
\end{eqnarray}
Fig.~\ref{Fig5} displays the integrated gamma-ray flux for two benchmark DM lifetimes, $7.3 \times 10^{28}$~s (red) and $2.9 \times 10^{30}$~s (orange), which correspond to the lower and upper bounds of the range allowed by the neutrino data, consistent with both the KM3-230213A signal and the IceCube-EHE upper limit. For comparison, the corresponding fluxes without attenuation are shown as dashed lines. Upper limits from various experiments are indicated by colored triangles: EAS-MSU (green)~\cite{Fomin:2017ypo}, Pierre Auger Observatory (blue)~\cite{Castellina:2019huz}, CASA-MIA (brown)~\cite{CASA-MIA:1997tns}, KASCADE (magenta)~\cite{KASCADEGrande:2017vwf}, and KASCADE-Grande (cyan)~\cite{KASCADEGrande:2017vwf}.
The flux for allowable DM lifetimes generally lies below all experimental limits, except for the upper edge of the red curve, which intersects the KASCADE-Grande data point near $1.38 \times 10^{7}$~GeV. It is interesting that an accurate treatment of gamma-ray attenuation effect avoid the tension from the KASCADE-Grande data.

\section{Cosmic Strings and Gravitational Waves}
\label{Cosmic string and Gravitational waves}

The spontaneous symmetry breaking of $\UoneX$ in our model typically leads to the formation of cosmic strings in the early universe—one-dimensional topological defects that emerge when the vacuum manifold $\mathcal{M}$ possesses a non-trivial first homotopy group, $\pi_1(\mathcal{M}) \neq 0$~\cite{Hindmarsh:1994re, Vachaspati:2015cma, 2000csot.book.....V}. Following their formation, the cosmic string network undergoes a transient evolution before reaching a scaling regime —an attractor solution characterized by the correlation length $L$ scaling linearly with cosmic time, $L \propto t$~\cite{1985NuPhB.252..227K, PhysRevLett.60.257, PhysRevLett.63.2776, PhysRevD.40.973, PhysRevLett.64.119}. Within this scaling regime, the energy density of long strings evolves as $\rho_\infty = \mu/L^2 \propto t^{-2}$, where $\mu$ denotes the tension of the string. This scaling behavior ensures that cosmic strings never dominate the universe's energy budget, as their energy density redshifts in the same manner as the dominant cosmic component: like radiation during the radiation-dominated era ($\rho \propto a^{-4}$) and like matter during the matter-dominated era ($\rho \propto a^{-3}$).

During the scaling regime, long cosmic strings frequently intersect and reconnect, producing loops that continuously drain energy from the long-string network. The subsequent oscillations and decay of these loops generate a stochastic gravitational wave background (SGWB) that persists to the present day. This SGWB spans a broad frequency range, making it an attractive target for next-generation GW interferometers.
In the following calculations for the SGWB of our model, we adopt the approach described in Ref.~\cite{Gouttenoire:2019kij}.

When a global or local $\Uone$ symmetry is spontaneously broken, the resulting cosmic strings are classified as global or local strings, respectively. For global strings, the dominant decay channel for loops is the emission of massless Goldstone bosons~\cite{2000csot.book.....V, Saurabh:2020pqe}, causing them to disappear after only a few oscillations and produce negligible GWs—except in scenarios where the VEV of scalar field reaches $\eta \gtrsim 10^{14}$~GeV~\cite{Chang:2019mza}. In contrast, the dominant energy loss for local strings is via GW emission~\footnote{Some studies suggest that local strings lose energy primarily through massive radiation, which can also maintain the scaling regime. See Refs.~\cite{Vincent:1997cx, Hindmarsh:2008dw, Hindmarsh:2017qff}.}, with the radiated power given by~\cite{2000csot.book.....V},
\begin{eqnarray}
P_\text{GW} = \Gamma G \mu^2,
\end{eqnarray}
where $\Gamma = \sum_k \Gamma^{(k)} \simeq 50$ is the total GW emission efficiency determined from Nambu-Goto simulations~\cite{Blanco-Pillado:2017oxo}, and $G$ is Newton's gravitational constant. Consequently, a loop formed at time $t_i$ with initial length $\alpha t_i$ shrinks due to GW radiation at a rate $\Gamma G \mu$, with its length at a later time $\tilde{t}$ given by,
\begin{eqnarray}
l(\tilde{t}) = \alpha t_i - \Gamma G \mu (\tilde{t} - t_i),
\end{eqnarray}
where $\alpha = 0.1$ represents the fraction of the horizon size—the characteristic scale at which GWs are predominantly produced. The GWs emitted by these loops have frequencies: $\tilde{f} = 2k/l(\tilde{t})$, where $k \in \mathbb{Z}^+$ denotes the mode number. The frequency observed today, at time $t_0$, is then redshifted as,
\begin{eqnarray}
f = \frac{a(\tilde{t})}{a(t_0)} \tilde{f}.
\end{eqnarray}

The GW power spectrum for the $k$-th mode, $P_\text{GW}^{(k)}$, depends on the small-scale structures in string loops, such as kinks and cusps~\cite{Damour:2001bk, Ringeval:2017eww}, and is given by,
\begin{eqnarray}
P_\text{GW}^{(k)} = \Gamma^{(k)} G \mu^2,~\text{with}\quad \Gamma^{(k)} = \frac{\Gamma k^{-n}}{\sum_{p=1}^\infty p^{-n}},
\end{eqnarray}
where $n = 4/3$, $5/3$, and $2$ for loops dominated by cusps, kinks, and kink-kink collisions, respectively. We adopt the cusp-dominated case ($n = 4/3$) in this work. The total SGWB spectrum from cosmic strings observed today is obtained by summing over all modes~\cite{Gouttenoire:2019kij}:
\begin{eqnarray}
\Omega_\text{GW}(f) \equiv \frac{f}{\rho_c} \frac{d \rho_\text{GW}}{df} = \sum_k \Omega_\text{GW}^{(k)}(f),
\end{eqnarray}
where
\begin{eqnarray}
\Omega_\text{GW}^{(k)}(f) = \frac{1}{\rho_c} \frac{2k}{f} \frac{\mathcal{F}_\alpha \Gamma^{(k)} G \mu^2}{\alpha(\alpha + \Gamma G \mu)} \int_{t_F}^{t_0} d\tilde{t} \frac{C_\text{eff}(t_i)}{t_i^4} \left( \frac{a(\tilde{t})}{a(t_0)}  \right)^5 \left( \frac{a(t_i)}{a(\tilde{t})}  \right)^3 \Theta(t_i - t_F).
\end{eqnarray}
Here, $\mathcal{F}_\alpha = 0.1$ represents the fraction of energy transferred from the string network into loops of size $\alpha$~\cite{Blanco-Pillado:2013qja}. We adopt the loop formation efficiency $C_\text{eff}$ to be $5.7$ during the radiation-dominated era and $0.5$ during the matter-dominated era—values derived from numerical simulations within the velocity-dependent one-scale (VOS) model~\cite{Cui:2017ufi, Blasi:2020wpy}. Additionally, $\Theta$ denotes the Heaviside function, and $t_F$ is the string network formation time, defined by: $\sqrt{\rho_\text{tot}(t_F)} \equiv \mu$, where $\rho_\text{tot}$ is the universe's total energy density. The loop formation time corresponding to mode $k$ is obtained through:
\begin{eqnarray}
t_i(f,\tilde{t}) = \frac{1}{\alpha + \Gamma G \mu} \left[ \frac{2k}{f} \frac{a(\tilde{t})}{a(t_0)} +  \Gamma G \mu \tilde{t}   \right].
\end{eqnarray}
It should be noted that we neglect contributions from thermal friction, arising from string interactions with plasma particles, as well as emissions of massive particles from small-scale structures like kinks and cusps. These effects mainly introduce cut-offs in the GW spectrum at ultra-high frequencies, which lie beyond the sensitivity range of current and upcoming GW interferometers~\cite{Gouttenoire:2019kij}.

\begin{figure}[h!]
\centering
\subfigure[\label{Fig6-1}]
{\includegraphics[width=0.55\textwidth]{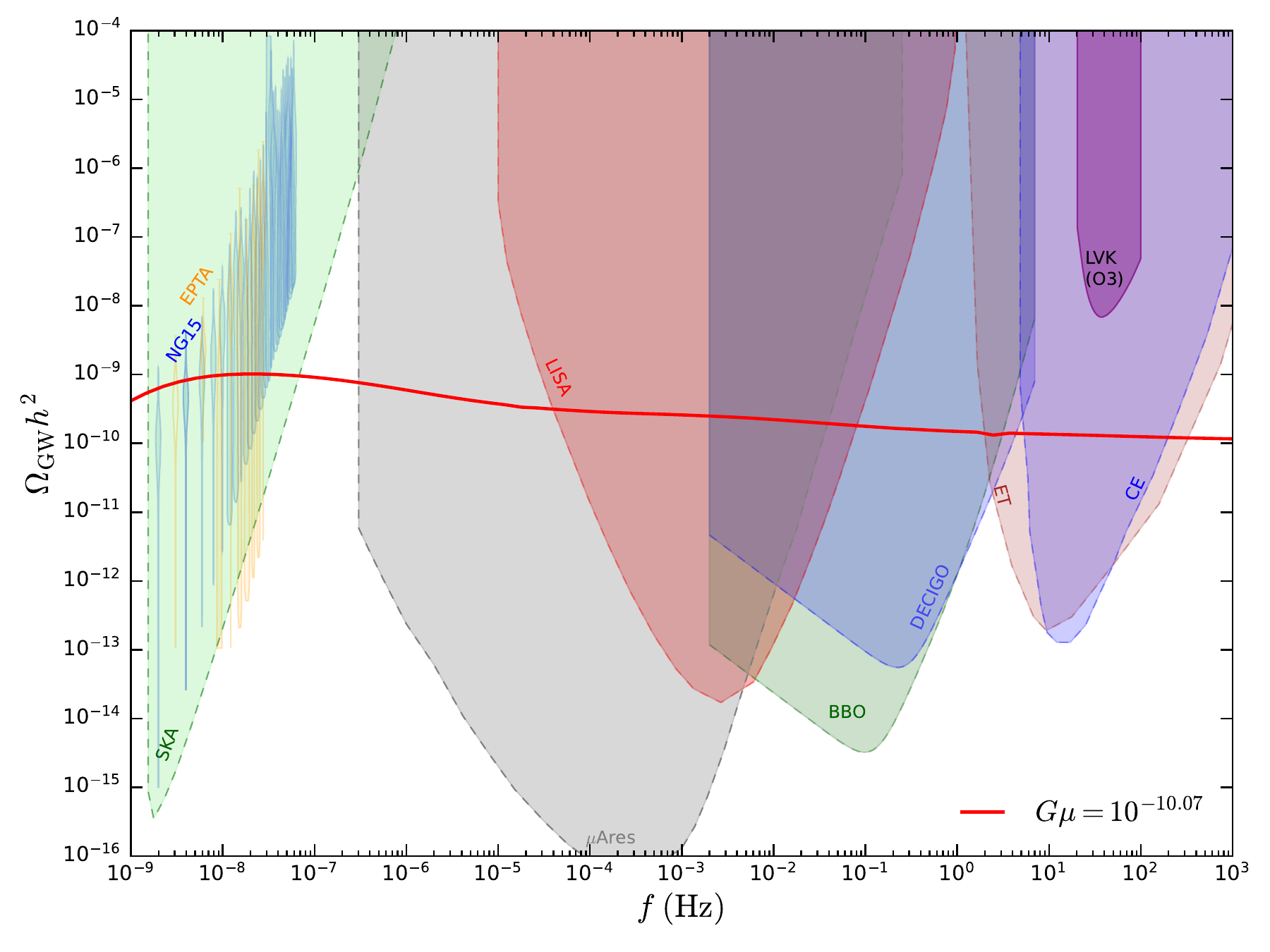}}
\hspace{.01\textwidth}
\subfigure[\label{Fig6-2}]
{\includegraphics[width=0.48\textwidth]{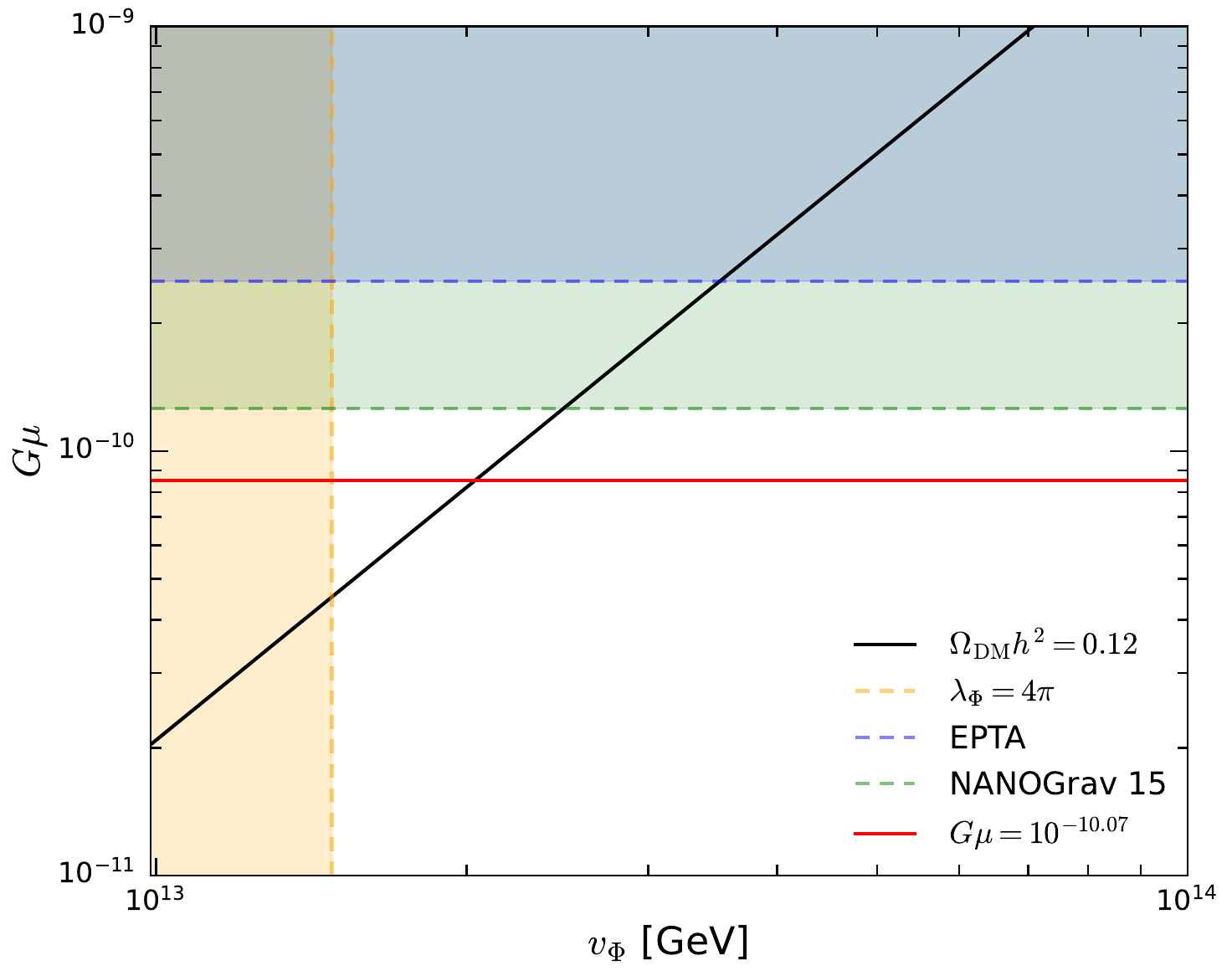}}
\hspace{.01\textwidth}
\subfigure[\label{Fig6-3}]
{\includegraphics[width=0.48\textwidth]{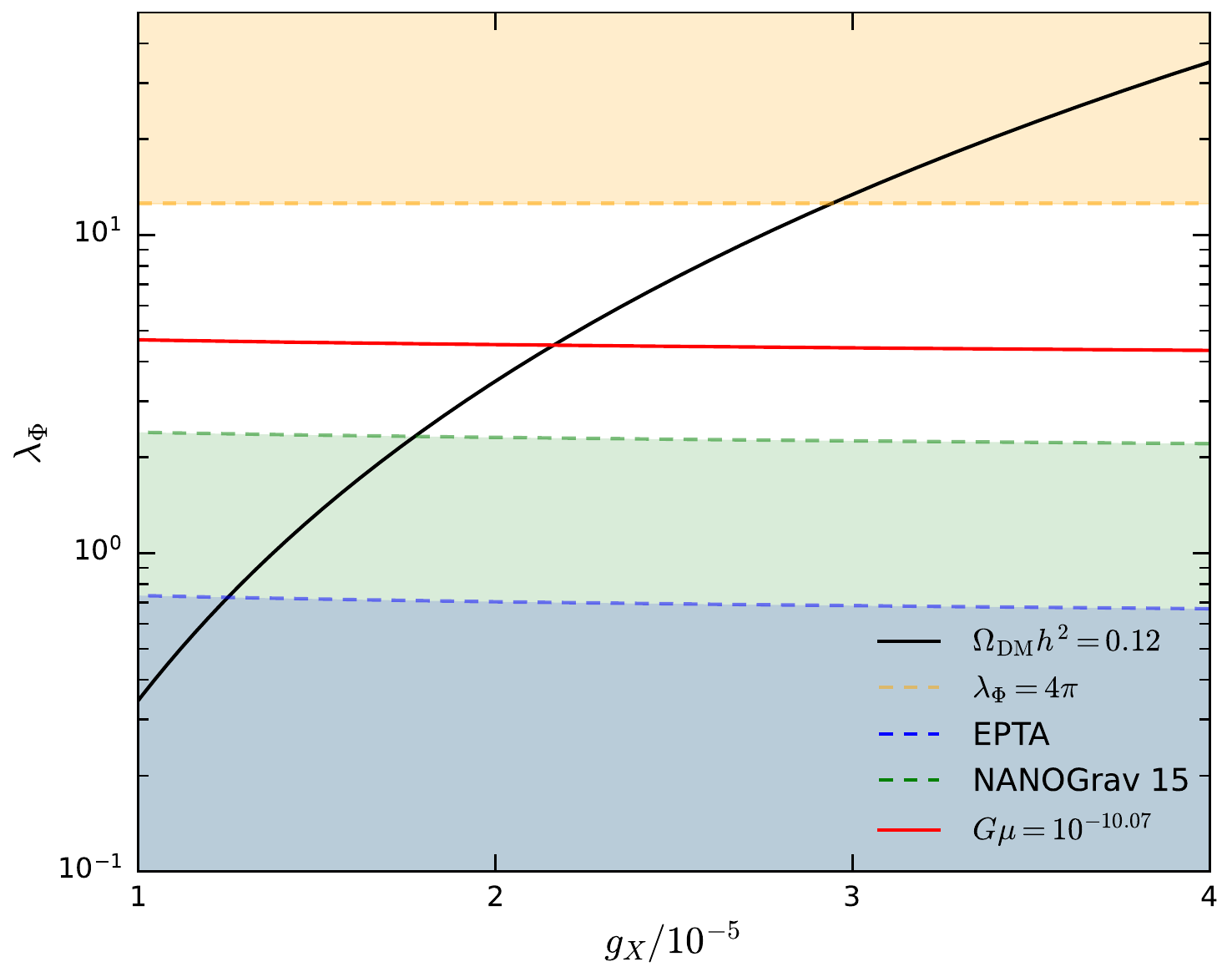}}
\caption{GW signatures from cosmic strings in our model. (a) Predicted GW spectrum for $G\mu = 10^{-10.07}$ compared with sensitivity curves from current and future detectors. (b) String tension $G\mu$ versus $v_\Phi$. (c) Scalar coupling $\lambda_\Phi$ versus gauge coupling $g_X$. In panels (b) and (c), the black curve represents parameter values yielding the correct DM relic density, the red curve corresponds to $G\mu = 10^{-10.07}$, the orange region indicates the parameter space excluded by the unitarity bound, and the blue/green curves show the $95\%$ confidence level upper limits from EPTA and NANOGrav observations, respectively.}
\label{Fig6}
\end{figure}

In Fig.~\ref{Fig6-1}, we present the predicted GW spectrum produced by cosmic strings for $G\mu = 10^{-10.07}$, corresponding to the median value obtained from fitting the SGWB spectrum in the nHz frequency band reported by the EPTA collaboration~\cite{EPTA:2023xxk}. The spectrum calculations were performed using the publicly available code \texttt{CosmicStringGW}~\cite{Fu:2024rsm}. For comparison, Fig.~\ref{Fig6-1} includes the projected sensitivity curves from several proposed detectors, including SKA~\cite{Janssen:2014dka}, $\mu$-Ares~\cite{Sesana:2019vho}, LISA~\cite{2017arXiv170200786A}, BBO/DECIGO~\cite{Yagi:2011wg}, ET~\cite{Hild:2008ng}, and CE~\cite{LIGOScientific:2016wof}. Additionally, upper bounds from LVK~\cite{LIGOScientific:2022sts, KAGRA:2021kbb, Jiang:2022uxp}, and recent observational results from NANOGrav~\cite{NANOGrav:2023gor, NANOGrav:2023hvm} and EPTA~\cite{EPTA:2023sfo, EPTA:2023xxk} are also displayed. As can be seen, the predicted GW spectrum for $G\mu = 10^{-10.07}$ achieves good agreement with the EPTA observations, but falls below the NANOGrav measurements in the relatively high frequency range ($3 \times 10^{-8}$ to $10^{-7}$ Hz). This discrepancy arises because large $G\mu$ values produce a relatively flat GW spectrum across the PTA band
, indicating that contributions from other sources may need to be considered in this frequency interval.

To explore the connection between the GW spectrum and our theoretical model, we examine different regions of the parameter space in Figs.~\ref{Fig6-2} and \ref{Fig6-3}. We adopt the relation between string tension $G\mu$ and the gauge coupling $g_X$, scalar coupling $\lambda_\Phi$, and VEV $v_\Phi$ as derived in Ref.~\cite{PhysRevD.37.263} (see Fig.~1). Fig.~\ref{Fig6-2} shows the predicted string tension $G\mu$ as a function of $v_\Phi$, while Fig.~\ref{Fig6-3} depicts $\lambda_\Phi$ versus $g_X$. In both figures, the black curve corresponds to parameters yielding the correct DM relic density $\Omega_\text{DM} h^2 = 0.12$ (as discussed in Fig.~\ref{Fig2-1}) and the red curve denotes the case with $G\mu = 10^{-10.07}$. The orange region is excluded by the perturbative unitarity bound $\lambda_\Phi < 4\pi$. The blue and green lines represent the $95\%$ C.L. upper limits set on $G\mu$ from EPTA ($G\mu < 10^{-9.6}$)~\cite{EPTA:2023xxk} and NANOGrav ($G\mu < 10^{-9.9}$)~\cite{NANOGrav:2023hvm}, respectively. Our model predicts a string tension $G\mu$  in the range $4.5 \times 10^{-11} \lesssim G\mu \lesssim 1.2 \times 10^{-10}$, corresponding to $v_\Phi$ between $1.5 \times 10^{13}$ and $2.5 \times 10^{13}$ GeV, $g_X$ between $1.8 \times 10^{-5}$ and $2.9 \times 10^{-5}$, and $\lambda_\Phi > 2.3$. In particular, the central value $G\mu = 10^{-10.07}$ reported by EPTA aligns with model parameters: $g_X = 2.2 \times 10^{-5}$, $\lambda_\Phi = 4.4$, and $v_\Phi = 2 \times 10^{13}$ GeV.

In summary, our model successfully accounts for the KM3NeT event through the appropriate DM mass and accurately reproduces the observed DM relic density, while simultaneously producing cosmic strings and generated GWs that satisfy current constraints from EPTA, NANOGrav, and LVK. The predicted GW spectrum will be further probed by upcoming interferometers such as $\mu$-Ares, LISA, BBO/DECIGO, ET, and CE. Remarkably, all these achievements are realized with only three model parameters: $(g_X, \lambda_\Phi, v_\Phi)$, making the model both theoretically elegant and experimentally testable across multiple independent channels.

\section{Conclusions}
\label{Conclusions}

In this work, we develop a DM model based on a new $\UoneX$ gauge symmetry, with its gauge boson field serving as the DM candidate. The model naturally realizes the minimal seesaw mechanism for neutrino mass generation while simultaneously achieving the correct DM relic density through freeze-in production via the heavy scalar portal. When the dark Higgs $\Phi$ acquires a VEV $v_\Phi$, the $\UoneX$ local symmetry is spontaneously broken, giving mass to the dark gauge boson. Moreover, the Yukawa interactions induce mixing between the Weyl fermions and the right-handed neutrinos, opening the channel for DM decay into light neutrinos.

To explain the KM3-230213A event, we calculate the predicted neutrino flux from our model, including both galactic and extragalactic contributions. Our analysis demonstrates that DM lifetimes in the range of $7.3 \times 10^{28}$~s to $2.9 \times 10^{30}$~s can successfully account for the observed UHE neutrino event while remaining consistent with existing experimental upper limits. We further examine the gamma-ray constraints by calculating the integral flux and comparing it with multiple experimental limits. This requires careful treatment of gamma-ray attenuation due to absorption by the ISRF during propagation from their production sites to Earth, particularly in the energy range of $10^5$~GeV to $10^9$~GeV. Our analysis shows that when absorption by the ISRF is properly included, the DM lifetime range required to explain the KM3-230213A event satisfies constraints from EAS-MSU, Pierre Auger Observatory, CASA-MIA, KASCADE, and KASCADE-Grande. However, neglecting the absorption effects would cause a portion of the parameter space to be excluded by KASCADE-Grande. Thus, the accurate treatment of gamma-ray attenuation significantly improves the robustness and viability of our results.

Additionally, the spontaneous breaking of $\UoneX$ naturally predicts the formation of cosmic strings in the early universe, resulting in an SGWB that can be tested by current and future GW observatories. Our model predicts a string tension in the range $4.5 \times 10^{-11} \lesssim G\mu \lesssim 1.2 \times 10^{-10}$, which greatly aligns with recent pulsar timing array observations. The central value $G\mu = 10^{-10.07}$ reported by EPTA corresponds to model parameters $g_X = 2.2 \times 10^{-5}$, $\lambda_\Phi = 4.4$, and $v_\Phi = 2 \times 10^{13}$ GeV, which simultaneously achieve the correct DM relic density and explain the KM3-230213A event. This multi-messenger consistency across UHE neutrinos, gamma-rays, and GW signatures validates our vector DM decay interpretation of the KM3-230213A event. The predicted GW spectrum offers promising detection prospects for upcoming interferometers, including $\mu$-Ares, LISA, BBO/DECIGO, ET, and CE, providing independent verification of the underlying $\UoneX$ symmetry breaking.

\begin{comment}
The detection of the KM3-230213A event by KM3NeT has invigorated the search for the origins of the most energetic cosmic neutrinos. While traditional astrophysical sources remain plausible, the hypothesis of super-heavy dark matter decay provides an attractive alternative that is testable through multi-messenger observations. Ongoing and future measurements by neutrino telescopes and gamma-ray observatories will be crucial in probing this scenario and potentially uncovering the fundamental nature of dark matter.
\end{comment}

\begin{acknowledgments}
We thank Zhao-Huan Yu for helpful discussions. C. C. is supported by the National Natural Science Foundation of China (NSFC) under Grant No. 11905300, and the Guangzhou Science and Technology Planning Project under Grant No. 2023A04J0008. H.-H. Z. is supported by the NSFC under Grant No. 12275367. This work is also supported by the
Fundamental Research Funds for the Central Universities, and the Sun Yat-Sen University Science Foundation.

\end{acknowledgments}

%%%%%%%%%%%%%%%
\bibliographystyle{utphys}
\bibliography{ref}

\end{document}